\documentclass[aps,pra,twocolumn,showpacs,superscriptaddress,longbibliography]{revtex4-2}
\usepackage{graphicx} 
\usepackage{amsmath}
\usepackage{graphicx,epstopdf}
\usepackage{gensymb}
\newcommand{\be}{\begin{equation}}
	\newcommand{\ee}{\end{equation}}
\newcommand{\bea}{\begin{eqnarray}}
	\newcommand{\eea}{\end{eqnarray}}
\newcommand{\bse}{\begin{subequations}}
	\newcommand{\ese}{\end{subequations}}

\usepackage{color}
\usepackage[colorlinks,bookmarks=false,citecolor=darkblue,linkcolor=red,urlcolor=blue]{hyperref}

\definecolor{darkred}{rgb}{0.7,0.0,0.0}

\definecolor{darkblue}{rgb}{0,0.02,0.45}

\definecolor{darkgreen}{rgb}{0.02,0.45,0.0}

\definecolor{violet}{rgb}{0.8,0.2,0.6}

\begin{document}

\title{Quasi-one-dimensional uniform spin-$\frac{1}{2}$ Heisenberg antiferromagnet KNaCuP$_2$O$_7$ probed by $^{31}$P and $^{23}$Na NMR}

\author{S. Guchhait}
\affiliation{School of Physics, Indian Institute of Science Education and Research Thiruvananthapuram-695551, India}
\author{Qing-Ping Ding}
\affiliation{Ames Laboratory and Department of Physics and Astronomy, Iowa State University, Ames, Iowa 50011, USA}
\author{M. Sahoo}
\affiliation{Department of Physics, University of Kerala, Kariavattom, Thiruvananthapuram-695581, India}
\author{A. Giri}
\author{S. Maji}
\affiliation{School of Physical Sciences, Indian Association for the Cultivation of Science, Kolkata-700032, India}
\author{Y. Furukawa}
\affiliation{Ames Laboratory and Department of Physics and Astronomy, Iowa State University, Ames, Iowa 50011, USA}
\author{R. Nath}
\email{rnath@iisertvm.ac.in}
\affiliation{School of Physics, Indian Institute of Science Education and Research Thiruvananthapuram-695551, India}
\date{\today}

\begin{abstract}
We present the structural and magnetic properties of KNaCuP$_2$O$_7$ investigated via x-ray diffraction, magnetization, specific heat, and $^{31}$P NMR and $^{23}$Na NMR measurements and complementary electronic structure calculations. The temperature dependent magnetic susceptibility and $^{31}$P NMR shift could be modeled very well by the uniform spin-$1/2$ Heisenberg antiferromagnetic chain model with nearest-neighbour interaction $J/k_{\rm B}\simeq 58.7$~K. The corresponding mapping using first principles electronic structure calculations leads to $J^{\rm DFT}/k_{\rm B} \simeq 59$~K with negligibly small inter-chain couplings, further confirming that the system is indeed an one-dimensional uniform spin-$1/2$ Heisenberg antiferromagnet. The diverging trend of NMR spin-lattice relaxation rates ($^{31}1/T_1$ and $^{23}1/T_1$) imply the onset of a magnetic long-range-ordering at around $T_{\rm N} \simeq 1$~K. From the value of $T_{\rm N}$, the average inter-chain coupling is estimated to be $J^{\prime}/k_{\rm B} \simeq 0.28$~K. Moreover, the NMR spin-lattice relaxation rates show the dominant contributions from uniform ($q=0$) and staggered ($q = \pm \pi/a$) spin fluctuations in the high and low temperature regimes, respectively mimicking one-dimensionality of the spin-lattice. We have also demonstrated that $^{31}1/T_1$ in high temperatures varies linearly with $1/\sqrt{H}$ reflecting the effect of spin diffusion on the dynamic susceptibility. The temperature-dependent unit cell volume could be described well using the Debye approximation with a Debye temperature of $\Theta_{\rm D} \simeq 294$~K, consistent with the heat capacity data.
\end{abstract}

\maketitle

\section{\textbf{Introduction}}
Quantum fluctuations play a pivotal role in deciding the ground state properties in low-dimensional spin systems~\cite{Mikeska2004,Parkinson2010}.
In particular, in uniform one-dimensional (1D) spin-$\frac{1}{2}$ Heisenberg antiferromagnetic (HAF) chains, quantum fluctuations are enhanced due to low spin value and reduced dimensionality which preclude magnetic LRO.\cite{Mermin1133} Often, the inter-chain and/or intra-chain frustration amplify the effect of quantum fluctuations leading to various intriguing low temperature features. Further, spin chains are the simplest systems which can be easily tractable from both experimental and computational point of views as they have relatively simple and well defined Heisenberg Hamiltonian $H = J\sum_{i} S_{i} S_{i+1}$, where $S_{i}$ and $S_{i+1}$ are the nearest-neighbour (NN) spins and $J$ is the exchange coupling between them. Transition metal oxides offer ample opportunities for finding spin chains with different exchange geometries. 

The copper (Cu$^{2+}$) based oxides are proven to be excellent model compounds and are extensively studied because of their interesting crystal lattice and low spin ($3d^9$, $S = 1/2$) value. The Cu$^{2+}$ chains formed by the direct linkage of CuO$_4$ units can be categorized into two groups. One is the chains formed by the edge-sharing of CuO$_4$ units and another formed by the corner sharing of CuO$_4$ units.
The chains of edge-sharing CuO$_4$ units have Cu-O-Cu angle nearly $90^{\degree}$ and are having competing NN ($J_1$) and next-nearest-neighbour (NNN) ($J_2$) interactions~\cite{Mizuno5326}. For AF $J_2$, these chains are frustrated, irrespective of the sign of $J_1$ and host a wide variety of ground states, controlled by the $J_2/J_1$ ratio.\cite{Furukawa257205} Prominent manifestation of frustration in 1D spin-$1/2$ chains encompasses spin-Peierls transition in CuGeO$_{3}$,\cite{Hase3651} chiral state in NaCu$_2$O$_2$,\cite{Caponaga140402} LiCu$_2$O$_2$,\cite{Masuda177201} LiCuVO$_4$,\cite{Buttgen214421} and Li$_2$ZrCuO$_4$,\cite{Drechsler077202} and realization of Majumdar-Ghosh point in Cu$_3$(MoO$_4$)(OH)$_4$~\cite{Lebernegg035145}.
In these compounds, $J_1$ and $J_2$ are comparable in strength, which generates a strong frustration within the chain. On the contrary, in Sr$_2 $CuO$_3$, chains are formed by corner sharing of CuO$_4$ units and is an ideal realization of spin-$1/2$ uniform HAF chains.\cite{Ami5994,Motoyama3212,Rosner3402,Takigawa4612,Thurber247202,Motoyama3212} Because of the nearly $180^{\degree}$ Cu-O-Cu angle, the AF $J_1$ prevails over $J_2$, largely reducing the inchain frustration and making the chains uniform.

\begin{figure*}
	\includegraphics[scale = 0.85]{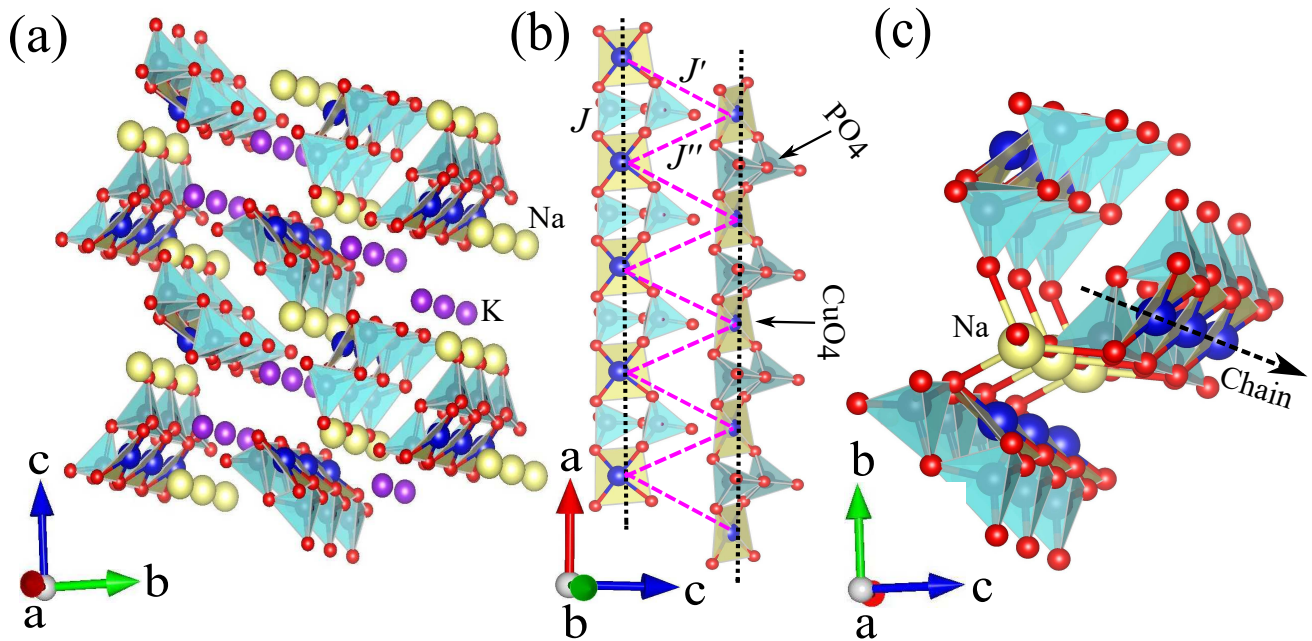} 
	\caption{(a) A three-dimensional view of the crystal structure of KNaCuP$_2$O$_7$ that shows well separated spin chains. (b) Two uniform spin chains of Cu$^{2+}$ running along the $a$-direction featuring the intrachain coupling ($J$) and the frustrated interchain network of $J^{\prime}$ [$d_{Cu-Cu} \simeq 5.772(2)$~\AA] and $J^{\prime \prime}$ [$d_{Cu-Cu} \simeq  5.676(2)$~\AA]. (c) A section of the crystal structure showing the coupling of Na atoms with Cu$^{2+}$ ions.}
	\label{Fig1}
\end{figure*}
Another family of 1D compounds is the copper phosphates (Sr,Ba)$_2$Cu(PO$_4$)$_2$, (Ba,Sr,Pb)CuP$_2 $O$_7$, and (Li,Na,K)$_2$CuP$_2 $O$_7$ which contain isolated CuO$_4$ units. \cite{Nath174436,Nath134451,Nath4285,Salunke085104,Lebernegg174436,Belik8572} Though there is no direct linking of CuO$_4$ units, the interaction among Cu$^{2+}$ ions takes place via an extended path involving the corner/edge sharing of CuO$_4$ and PO$_{4}$ tetrahedra.
The magnetic properties of all these compounds are described well by the 1D uniform spin-$1/2$ HAF model with intra-chain coupling $J/k_{\rm B} (=J_1/k_{\rm B})$ in the range $\sim 30$~K to 160~K. (Sr,Ba)$_2$Cu(PO$_4)_2$ are emerged to be best realization of uniform spin-$1/2$ HAF chains showing one-dimensionality over a large temperature range ($k_{\rm B}T/J \geq 6 \times 10^{-4}$), similar to Sr$_2$CuO$_3$ ($k_{\rm B}T/J \geq 2 \times 10^{-3}$).\cite{Nath174436,Motoyama3212} Spin chains based on organo-metallic complexes are another class of compounds portraying interesting 1D physics~\cite{Hammar1008,*Hong132404}. When the spin chains are embedded in a real material, a weak residual coupling between the chains comes into play at sufficiently low temperatures and the ground state is decided based on the hierarchy of coupling strengths. These inter-chain couplings often form a frustrated network between the chains and either forbid the system in crossing over to a LRO state or stabilize in a exotic ground state~\cite{Ahmed214413}. Thus, the quest for novel states in spin chains necessitates the search for new model compounds with non-trivial inter-chain geometries.

Herein, we investigate the magnetic behavior of potassium sodium copper(II) diphosphate(V) (KNaCuP$_2$O$_7$), which has a monoclinic crystal structure with space group $P2_1/n$. The lattice parameters and unit cell volume $(V_{\rm cell})$ at room temperature are reported to be $a= 5.176(3)$~\AA, $b = 13.972(5)$~\AA, $c= 9.067(3)$~\AA, $\beta= 91.34(2)^\circ$, and $V_{\rm cell}= 655.6(5)$~\AA$^3$.\cite{Fitouri109} 
The crystal structure of KNaCuP$_2$O$_7$ is presented in Fig.~\ref{Fig1}. Distorted CuO$_4$ plaquettes are corner shared with four PO$_4$ tetrahedra forming isolated magnetic chains stretched along the $a$-direction.
In each CuO$_4$ plaquette, Cu-O bond lengths are within the range 1.93-1.98~\AA~while in each PO$_4$ tetrahedra, the P-O bond length varies within the range 1.48-1.63~\AA. These chains are well separated from each other and the Na and K atoms are located in the interstitial positions between the chains. Thus, P is located almost symmetrically between two Cu$^{2+}$ ions within a chain and is strongly coupled with the magnetic Cu$^{2+}$ ions. The Na and K atoms are also positioned symmetrically between the chains, providing a weak interchain coupling and making a complex three-dimensional (3D) structure. Further, the chains are arranged in such a way that each CuO$_4$ plaquette in one chain has two identical neighbours in each adjacent chain. With AF $J$, $J^{\prime}$, and $J^{\prime\prime}$ this leads to a frustrated interchain geometry. Figure~\ref{Fig1}(b) presents a sketch of the spin lattice illustrating the leading intrachain ($J$) and the frustrated interchain couplings ($J^{\prime}$, $J^{\prime\prime}$) between two neighbouring chains. Moreover, only one Cu site in the crystal structure and the presence of inversion centers in the middle of each Cu–Cu bond imply that the anisotropic Dzyaloshinskii-Moriya (DM) interaction vanishes by symmetry.
Figure~\ref{Fig1}(c) shows a section of the crystal structure demonstrating the coupling of Na atom with three neighbouring chains. The magnetic properties of this compound are not available to date.


Our experimental results reveal uniform spin-$1/2$ chain character of the spin-lattice with a intra-chain coupling $J/k_{\rm B} \simeq 58.7$~K. The magnetic LRO is suppressed to $T_{\rm N} \simeq 1$~K due to weak and frustrated inter-chain couplings. The experimental assessment of the spin-lattice is further supported by the complementary electronic structure calculations. The dynamical properties of the spin system are also extensively investigated via $^{31}$P and $^{23}$Na NMR spin-lattice relaxation measurements.

\section{Methods}
\begin{figure}
	\includegraphics[width=\linewidth]{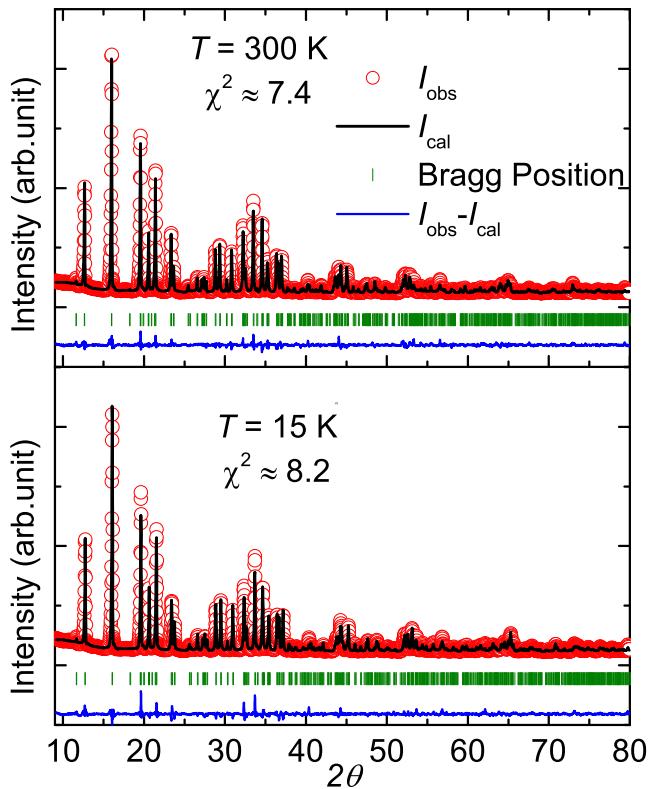}
	\caption{Powder XRD patterns (open circles) at room temperature (300~K) and 15~K for KNaCuP$_2$O$_7$. The solid line is the Rietveld fit, the vertical bars mark the expected Bragg peak positions, and the lower solid line corresponds to the difference between the observed and calculated intensities.}
	\label{Fig2}
\end{figure}
Blue colored polycrystalline sample of KNaCuP$_2$O$_7$ was synthesized by the traditional solid state synthesis procedure. Stoichiometric amount of CuO (Aldrich, 99.999$\%$), NaH$_4$PO$_5$ (Aldrich, 98$\%$), and KHPO$ _4$ were ground thoroughly and heated at 450$^\circ$C for 24~hrs in air. Subsequently, the sample was fired at 570$^\circ$C for 24~hrs and at 600$^\circ$C for 48~hrs followed by intermediate grindings and palletizations. Finally, the main phase was found to be formed at 600$^\circ$C. At each step, the phase purity of the sample was checked by doing powder x-ray diffraction (XRD) experiment at room temperature using a PANalytical powder diffractometer equipped with Cu~$K_{\alpha}$ radiation ($\lambda_{\rm avg} \simeq 1.54182$~\AA).
The temperature ($T$) dependent powder XRD was performed on the phase pure sample in the temperature range 15~K~$\leq T\leq300$~K, using a low temperature attachment (Oxford Phenix) to the x-ray diffractometer. Rietveld analysis of the XRD patterns was performed using the FULLPROF software package \cite{Carvajal55}, taking the initial structural parameters from Ref.~\cite{Fitouri109}.

Magnetization ($M$) was measured as a function of temperature (2~K~$\leq T\leq350$~K), in the presence of an applied magnetic field $H = 1$~T. Magnetization isotherms ($M$ vs $H$) were also measured at two different temperatures ($T = 2$~K, and 300~K) by varying $H$ from 0 to 9~T. All these measurements were carried out using a vibrating sample magnetometer (VSM) attachment to the Physical Property Measurement System (PPMS, Quantum Design). Specific heat ($C_{\rm p }$) was measured as a function of temperature (2 - 100~K), by using the thermal relaxation method in PPMS, on a sintered pellet in zero magnetic field. Magnetic spin susceptibility of a uniform AF chain lattice of Heisenberg spins was obtained from the quantum Monte Carlo (QMC) simulations performed with the LOOP algorithm \cite{Todo047203} of the ALPS simulation package \cite{BauerP05001}. Simulations were performed on a finite lattice ($L=200$) size.

The pulsed NMR experiments were performed on the $^{31}$P nucleus with nuclear spin $I = \frac{1}{2}$ and gyromagnetic ratio $\frac{\gamma}{2\pi} =17.237$~MHz/T and $^{23}$Na nucleus with $I=3/2$ and $\frac{\gamma}{2\pi} = 11.26$~MHz/T. $^{31}$P NMR measurements were done in different radio frequencies of 121~MHz, 85~MHz, 39~MHz, 21~MHz, and 11.6~MHz while $^{23}$Na NMR measurements were done in 79~MHz. The NMR spectrum at different temperatures was obtained by changing the magnetic field in a fixed frequency. A large temperature range of 1.6~K~$\leq T \leq 300$~K was covered in our experiments. Temperature dependent NMR shift $K(T)= [H_{\rm ref}/H(T) - 1]$ was calculated from the resonance field of the sample $H$ with respect to the resonance field of a non-magnetic reference sample $(H_ {\rm ref})$. The spin-lattice relaxation rate $1/T_1$ was measured by the conventional single saturation pulse method.

The first principles electronic structure calculations have been performed within the framework of density functional theory (DFT) using the plane-wave basis with projector augmented wave (PAW) potential\cite{Blochl17953,Kresse1758} as implemented in the Vienna Abinitio Simulation Package (VASP)\cite{Kresse558,Kresse11169}. Generalized gradient approximation (GGA) implemented within Perdew-Burke-Ernzerhof (PBE) prescription\cite{Perdew3865} has been chosen for the exchange-correlation functional. A plane wave cut-off of 500~eV was set to obtain good convergence of total energy and a k-mesh of $5 \times 2 \times 3$ was used for the Brilliouin zone (BZ) integration. Maximally localized Wannier functions (MLWFs) for the low energy Cu~d$_{x^{2}-y^{2}}$ model Hamiltonian have been constructed using VASP2WANNIER and WANNIER90 codes\cite{Mostofi2309}, providing the hopping parameters required to identify the various exchange paths. The missing correlation in GGA calculations are included within GGA+U method for all the spin-polarised calculations, where standard values of $U$ and Hund's coupling $J_{\rm H}$\cite{Bhowal075110} was chosen for Cu with $U_{\rm eff} (= U-J_{\rm H}) = 6.5$~eV in the Dudarev's scheme\cite{Dudarev1505}.

\section{Results}
\subsection{X-ray Diffraction}
\begin{figure}
	\includegraphics[width=\linewidth]{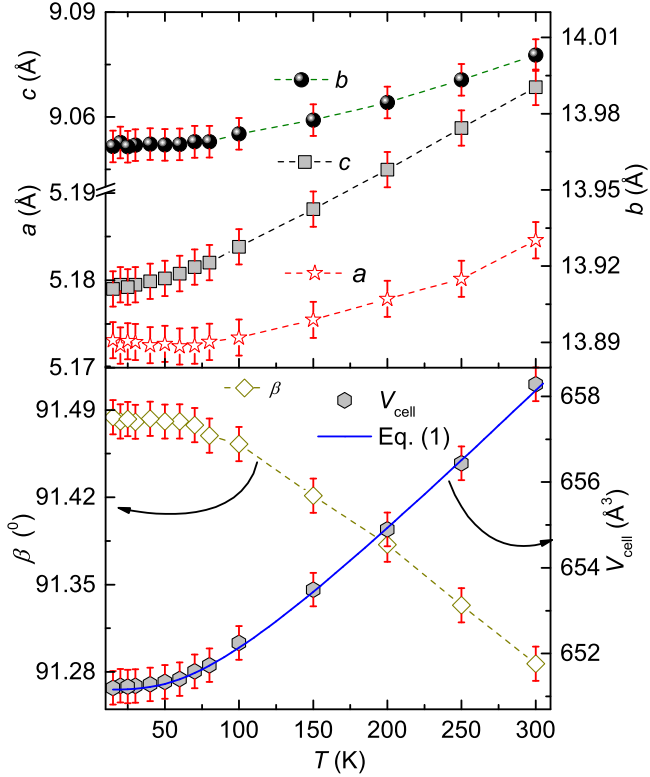}
	\caption{The lattice constants ($a$, $b$, and $c$), monoclinic angle ($\beta$), and unit cell volume ($V_{\rm cell}$) are plotted as a function of temperature from 15~K to 300~K. The solid line in the bottom panel represents the fit using Eq.~\eqref{Vcell}.}
	\label{Fig3}
\end{figure}
The powder XRD patterns of KNaCuP$_2$O$_7$ along with the Rietveld refinement are shown in Fig.~\ref{Fig2} for two different temperatures ($T$ = 300~K and 15~K). All the XRD patterns down to 15~K could be refined using the same crystal structure (Monoclinic, Space Group: $P2_1/n$), which indicates that there is neither any structural transition nor lattice distortion. The appearance of sharp and high intensity peaks with no extra reflections further reflects high quality and phase pure sample. From the refinement, the goodness-of-fit is achieved to be $\chi^{2}\sim 7.4$ and $\sim 8.2$ for $T = 300$~K and $15$~K, respectively. The refined lattice parameters and unit cell volume are [$a= 5.1846(1)$~\AA, $b=13.9904(2)$~\AA, $c=9.0777(2)$~\AA, $\beta= 91.286(2)^\circ$, and $V_{\rm cell}\simeq658.281$~\AA$^3$] and [$a= 5.1731(1)$~\AA, $b=13.9110(2)$~\AA, $c=9.0515(1)$~\AA, $\beta= 91.484(2)^\circ$, and  $V_{\rm cell}\simeq651.20$~\AA$^3$] for $T = 300$~K and 15~K, respectively. The refined structural parameters at room temperature are in close agreement with the values reported earlier.\cite{Fitouri109} Moreover, $V_{\rm cell} \simeq658.281$~\AA$^3$ at room temperature is found to have an intermediate value between K$_2$CuP$_2$O$_7$ ($\sim 721.88$~\AA$^3$), Li$_2$CuP$_2$O$_7$ ($\sim 585.24$~\AA$^3$), and Na$_2$CuP$_2$O$_7$ ($\sim 612.88$~\AA$^3$), as expected based on the ionic radii of K$^{1+}$, Li$^{1+}$, and Na$^{1+}$~\cite{Erragh23,*Elmaadi13,*Gopalakrishna1171}. Hence, one may also expect the magnetic parameters of KNaCuP$_2$O$_7$ to have values between K$_2$CuP$_2$O$_7$ and (Li,Na)$_2$CuP$_2$O$_7$, as change in volume brings in a change in interatomic distances.
The obtained temperature dependent lattice parameters ($a$, $b$, $c$, and $\beta$) and unit cell volume ($V_{\rm cell}$) are plotted in Fig.~\ref{Fig3}. The lattice constants $a$, $b$, and $c$ are found to be decreasing in a systematic way, while monoclinic angle $\beta$ is increasing with decreasing temperature. These lead to a overall decrease of $V_{\rm cell}$ with temperature.

The variation of unit cell volume with temperature can be expressed in terms of the Grüneisen $(\gamma)$ ratio as $\gamma = V_{\rm cell}(\frac{\partial P}{\partial U})_{V_{\rm cell}}=\frac{\alpha V_{\rm cell} K_{0}}{C_{\rm v}}$, where $\alpha$ is the thermal expansion coefficient, $C_{\rm v}$ is the heat capacity at constant volume, $K_0$ is the bulk modulus, and $U(T)$ is the internal energy of the system.\cite{Budd18} Assuming both $\gamma$ and $K_{0}$ are independent of temperature, $V_{\rm cell}(T)$ can be written as\cite {Wallace1998}
\begin{equation}\label{Vcell}
V_{\rm cell}(T) = \frac{\gamma U(T)}{K_0} + V_0,
\end{equation}
where $V_0$ is the unit cell volume at $T = 0$~K. According to the Debye model, $U(T)$ can be written as
\begin{equation}\label{Uenergy}
U(T) = 9Nk_{\rm B}T\left(\frac{T}{\theta_{\rm D}}\right)^3 \int_{0}^{\frac{\theta_{\rm D}}{T}}\frac{x^3}{(e^{x}-1)}dx,
\end{equation}
where, $N$ is the number of atoms per unit cell, $k_{\rm B}$ is the Boltzmann constant, and $\theta_{\rm D}$ is the average Debye temperature~\cite{Kittel1986}. The variable $x$ inside the integration stands for the quantity $\frac{\hbar\omega}{k_{\rm B}T}$ with phonon frequency $\omega$ and Planck constant $\hbar$. 
The fit of the experimental $V_{\rm cell}(T)$ data by Eq.~\eqref{Vcell} is shown as a solid line in the lower panel of Fig.~\ref{Fig3}. The obtained best fit parameters are $\theta_{\rm D} \simeq 294$~K, $V_0\simeq 651.19$~\AA$^3$, and $\frac{\gamma}{K_0}\simeq 1.14 \times 10^{-4}$~Pa$^{-1}$.

\subsection{Magnetization} 
\begin{figure}
	\includegraphics[width=\linewidth]{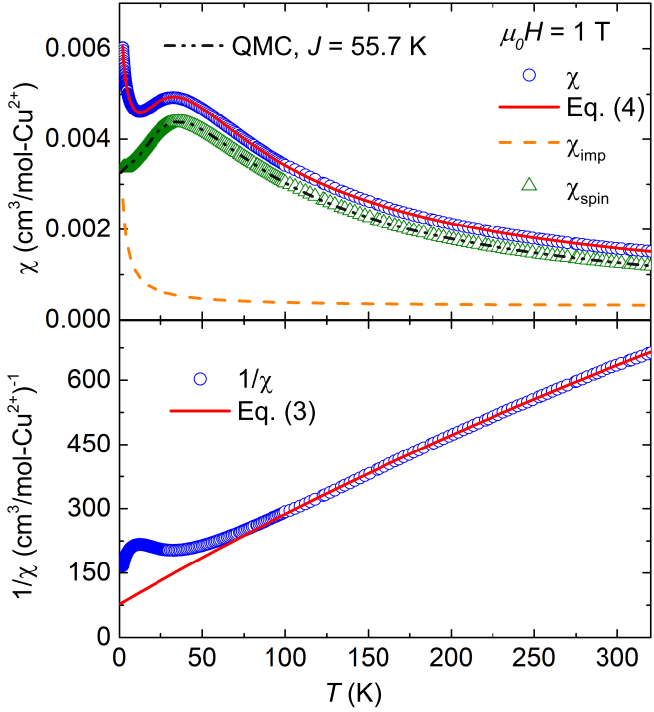}
	\caption{Upper panel: $\chi$ vs $T$ of KNaCuP$_2$O$_7$ in an applied field of 1~T and the red solid line is the best fit using Eq.~\eqref{chi_1D}. The dashed line represents the impurity contribution, $\chi_{\rm imp}(T) = \chi_0 +\frac{C_{\rm imp }}{T + \theta_{\rm imp}}$, obtained from the fit. The spin susceptibility $\chi_{\rm spin}(T)$ is obtained by subtracting $\chi_{\rm imp}(T)$ from $\chi(T)$. The dash-dotted line illustrates the QMC data with $J/k_{\rm B} = 55.7$~K and $g = 2.1$. Lower panel: Inverse magnetic susceptibility ($1/\chi$) as a function of $T$ and the solid line is the Curie-Weiss fit.}
	\label{Fig4}
\end{figure}
The magnetic susceptibility [$\chi(T) \equiv M/H$] of KNaCuP$_2$O$_7$ measured in an applied field $H = 1$~T is shown in the upper panel of Fig.~\ref{Fig4}. At high temperatures, $\chi(T)$ follows the standard paramagnetic behaviour and then passes through a broad maximum at around $T_{ \chi}^{\rm max}\simeq 35$~K. This broad maximum is a clear signature of the short-range ordering. At low temperatures, it shows a upturn which could be due to extrinsic paramagnetic impurities, defects, and/or uncorrelated spins at the open end of the finite chains in the powder sample~\cite {Fujimoto037206,Zvyagin214430}.
No indication of any magnetic LRO was found down to 2~K.

The inverse susceptibility, $1/\chi(T)$ is shown in the bottom panel of Fig.~\ref{Fig4}. The data in the paramagnetic regime are fitted by the Curie-Weiss (CW) law
\begin{equation}\label{cw}
\chi(T) = \chi_0 + \frac{C}{T + \theta_{\rm CW}}.
\end{equation}
Here, $\chi_0$ is the temperature-independent susceptibility, which includes Van-Vleck paramagnetic susceptibility (due to open electron shells of Cu$^{2+}$ ions) and core diamagnetic susceptibility (due to the core electron shells), $C$ is the Curie constant, and $\theta_{\rm CW}$ is the CW temperature. The fit in the temperature range $T\geq 100$~K yields the parameters: $\chi_0\simeq 2.01\times 10^{-4}$~cm$^3$/mol-Cu$^{2+}$, $C\simeq0.425$~cm$^3$K/mol-Cu$^{2+}$, and $\theta_{\rm CW}\simeq +33$~K. Using the value of $C$, the effective moment can be estimated as $\mu_{\rm eff} =(3k_{\rm B}C/N_{\rm A}\mu_{\rm B}^{2})^{\frac{1}{2}}$, where $N_{\rm A}$ is the Avogadro’s number and $\mu_{\rm B}$ is the Bohr magneton. Our experimental value of $C$ corresponds to $\mu_{\rm eff} \simeq 1.84\mu_{\rm B}$/Cu$^{2+}$. This value of $\mu_{\rm eff}$ is slightly greater than the ideal value $1.73\mu_{\rm B}$, for spin-$1/2$ and is typical for Cu$^{2+}$ based compounds.\cite{Janson094435,Nath014407} The positive value of $\theta_{\rm CW}$ indicates the AF exchange coupling between the Cu$^{2+}$ ions. The core diamagnetic susceptibility ($\chi_{\rm core}$) of the compound was calculated to be $-1.15\times10^{-4}$~cm$^3$/mol by adding the core diamagnetic susceptibility of Na$^{+}$, K$^{+}$, Cu$^{2+}$, P$^{5+}$, and O$^{2-}$ ions.\cite{selwood2013} The Van-Vleck paramagnetic susceptibility ($\chi_{\rm vv}$) was estimated to be $\sim 3.16\times10^{-4}$~cm$^3$/mol by subtracting $\chi_{\rm core}$ from $\chi_0$ which is very close to the value reported for other Cu$^{2+}$ based compounds.\cite{Islam174432,Nath174436,Motoyama3212}

In order to understand the spin-lattice, $\chi(T)$ was fitted by the uniform spin-$1/2$ Heisenberg chain model, taking into account the temperature independent ($\chi_0$) and extrinsic paramagnetic contributions. For the purpose of fitting, one can write $\chi(T)$ as the sum of three parts
\begin{equation}
\label{chi_1D}
\chi(T)=\chi_0+\frac{C_{\rm imp }}{T+\theta_{\rm imp}}+\chi_{\rm spin}(T).
\end{equation}
Here, the second term accounts for the paramagnetic impurity contributions with $\theta_{\rm imp}$ being the interaction strength between the impurity spins and $\chi_{\rm spin}(T)$ represents the spin susceptibility of a spin-$1/2$ uniform Heisenberg AF chain. We have used the expression of $\chi_{\rm spin}(T)$ given by Johnston~$et$~$al$.~\cite{Johnston9558}, which predicts the spin susceptibility accurately over a wide temperature range $5 \times 10^{-25} \leq k_{\rm B}T/J \leq 5$. Our experimental data in the whole measured temperature range were fitted well by Eq.~\eqref{chi_1D}, reflecting purely 1D character of the compound. As shown in Fig.~\ref{Fig4} (upper panel), the best fit yields the intra-chain coupling $J/k_{\rm B}\simeq 55.5$~K, $\chi_0\simeq 2 \times 10^{-4}$~cm$^3$/mol, $C_{\rm imp}\simeq0.0089$~cm$^3$K/mol, $\theta_{\rm imp}\simeq 1.74$~K, and Landé $g$-factor $g \simeq 2.1$. The value of $C_{\rm imp}$ corresponds to the impurity concentration of nearly $\sim 2.1$~\%, assuming impurity spins $S=\frac{1}{2}$.
A slightly larger value of $g$ $(\textgreater~2)$ is typically observed from electron-spin-resonance (ESR) experiments on Cu$^{2+}$ based compounds~\cite{Lebernegg174436}.

The intrinsic $\chi_{\rm spin}(T)$ of KNaCuP$_2$O$_7$ obtained after subtracting the temperature independent and paramagnetic impurity contributions from $\chi(T)$ is also shown in Fig.~\ref{Fig4}(upper panel). We also simulated $\chi_{\rm spin}(T)$ using QMC simulation considering a uniform chain model with $J/k_{\rm B} = 55.7$~K and $g = 2.1$ [see Fig.~\ref{Fig4}(upper panel)]. The simulated data without any additional term reproduce $\chi_{\rm spin}(T)$ perfectly in the whole temperature range. Indeed, our estimated quantities $\chi_{\rm spin}^{\rm max}J/N_{\rm A}g^{2}\mu_{\rm B}^{2} \simeq 0.1464$ and $\chi_{\rm spin}^{\rm max}T^{\rm max}_{\chi}/g^{2} \simeq 0.03512$~cm$^3$-K/mol (where $\chi_{\rm spin}^{\rm max}=0.00438$~cm$^3$/mol is the maximum in $\chi_{\rm spin}$ at $T_{\chi}^{\rm max}$ in Fig.~\ref{Fig4}) are quite consistent with the theoretically predicted values $\chi_{\rm spin}^{\rm max}J/N_{\rm A}g^{2}\mu_{\rm B}^{2}= 0.146926279$ and $\chi_{\rm spin}^{\rm max}T^{\rm max}_{ \chi}/g^2=0.0353229$~cm$^3$~K/mol~\cite{Klumper4701,Johnston9558}, endorsing the 1D spin-$1/2$ uniform HAF nature of the spin-lattice in KNaCuP$_2$O$_7$.


\begin{figure}
	\includegraphics[width=\linewidth]{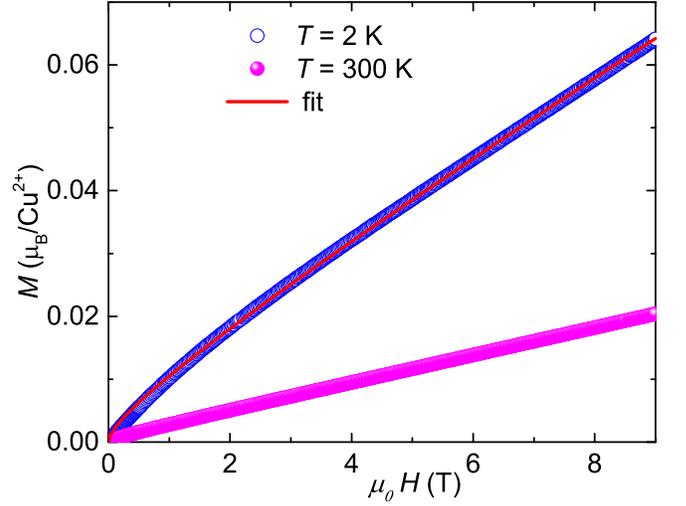}
	\caption {Magnetization $(M)$ of KNaCuP$_2$O$_7$ as a function of magnetic field $(H)$ at two different temperatures. The solid line is the fit to the magnetic isotherm at $T=2$~K, as described in the text.}
	\label{Fig5}
\end{figure}
Magnetization isotherm ($M$ vs $H$) measured at two end temperatures ($T=2$~K and 300~K) are shown in Fig.~\ref{Fig5}. For $T=300$~K, $M$ increase linearly with $H$, as expected for typical AFs at high temperatures. On the other hand, for $T = 2$~K, the behaviour is found to be non-linear and $M$ reaches a value $\sim 0.064$~$\mu_{\rm B}$/Cu$^{2+}$ at 9~T which is far below the saturation value $1 \mu_{\rm B}$. This is because, our maximum measured field of 9~T is far below the expected saturation field $H_{\rm s}=2J/g\mu_{\rm B} \simeq 78.5$~T, taking $J/k_{\rm B} \simeq 55.5$~K~\cite{Lebernegg174436}. Further, the magnetization data at $T = 2$~K were fitted well using the phenomenological expression for a spin chain, $M_{\rm chain} = \alpha H + \beta \sqrt{H}$. The obtained parameters $\alpha \simeq 5.46\times 10^{-7}$ and $\beta \simeq 5.02\times 10^{-5}$ are comparable with the values reported for spin-$1/2$ chain compound Bi$_6$V$_3$O$_{16}$~\cite{Chakrabarty094431}.

\subsection{Specific Heat}
\begin{figure}
	\includegraphics[width=\linewidth]{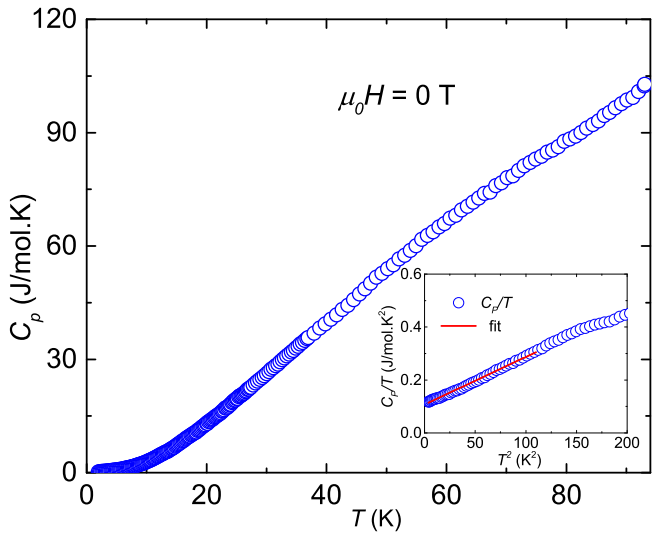}
	\caption {$C_{\rm p}$ of KNaCuP$_2$O$_7$ as a function of temperature in the absence of magnetic field. Inset: $C_p/T$ vs $T^2$ at low temperatures.}
	\label{Fig6}
\end{figure}
Temperature dependent specific heat $C_{\rm p}(T)$ measured in zero applied field is shown in Fig.~\ref{Fig6}. No anomaly associated with the magnetic LRO was noticed down to 2~K, consistent with the $\chi(T)$ data. In a magnetic insulator, there are two major contributions to the specific heat: phonon excitations and magnetic contribution. In the high-temperature region ($T > J/k_{\rm B}$), $C_{\rm p}$ is mainly dominated by phonon excitations, whereas the magnetic part contributes only in the low temperature region.

In the low temperature regime, $C_{\rm p}(T)$ can be fitted by $C_{\rm p}=\gamma T+\beta T^{3}$ where the cubic term accounts for the phononic contribution to the specific heat $(C_{\rm ph})$ and the linear term represents the magnetic contribution to the specific heat $(C_{\rm mag})$. In the inset of Fig.~\ref{Fig6}, $C_{\rm p}/T$ is plotted against $T^2$ which follows a linear behaviour in the low temperature regime. For a gapless spin-$1/2$ 1D HAF chain, $C_{\rm mag}(T)$ at low temperatures is expected to be linear with temperature and the linear coefficient ($\gamma$) provides a measure of $J/k_{\rm B}$. From the theoretical calculations, Johnston and Kl{\"u}mper have predicted the relation $\gamma_{\rm theory}=\frac{2R}{3(J/k_{\rm B})}$ for low temperatures $T < 0.2J/k_{\rm B}$~\cite{Klumper677,Johnston9558}. Using the value of $J/k_{\rm B}\simeq 55.5$~K, it is calculated to be $\gamma_{\rm theory}\simeq 0.1$~J/mol.K$^{2}$ for KNaCuP$_2$O$_7$.
The $C_{\rm p}/T$ vs $T^{2}$ data in the temperature range $T \leq 10$~K were fitted by the above equation and the extracted parameters are $\gamma_{\rm expt}\simeq 0.107$~J/mol.K$^{2}$ and $\beta\simeq 0.0018$~J/mol.K$^{4}$. The value of $\gamma_{\rm expt}$ is indeed very close to $\gamma_{\rm theory}$.
Following the Debye model, one can write $\beta=12\pi^{4}mR/5\theta_{\rm D}^{3}$ where $m$ is the total number of atoms in the formula unit and $R$ is the universal gas constant~\cite{Kittel1986}. From the value of $\beta$ the corresponding Debye temperature is estimated to be $\theta_{\rm D} \simeq 235$~K, which is close to the value obtained from the $V_{\rm cell}$ vs $T$ analysis~\footnote{Since $C_{\rm p} \simeq \frac{12\pi^{4}mR}{5} (\frac{T}{\theta_D})^3$ is the low temperature approximation of the Debye model, the value of $\theta_{\rm D}$ obtained from the low-$T$ $C_{\rm p}(T)$ data will always be less than the value obtained from the fit of the Debye model to the data in the whole temperature range.}.

\subsection{NMR}
NMR is an extremely powerful local tool to investigate the static and dynamic properties of a spin system. In KNaCuP$_2$O$_7$, P is coupled strongly while Na which is located inbetween the chains is coupled weakly to the Cu$^{2+}$ ions (see Fig.~\ref{Fig1}). Therefore, one can extract information about Cu$^{2+}$ spins by probing at the $^{31}$P and $^{23}$Na nuclear sites.

\subsubsection{$^{31}$P NMR Spectra}
\begin{figure}
	\includegraphics[width=\linewidth]{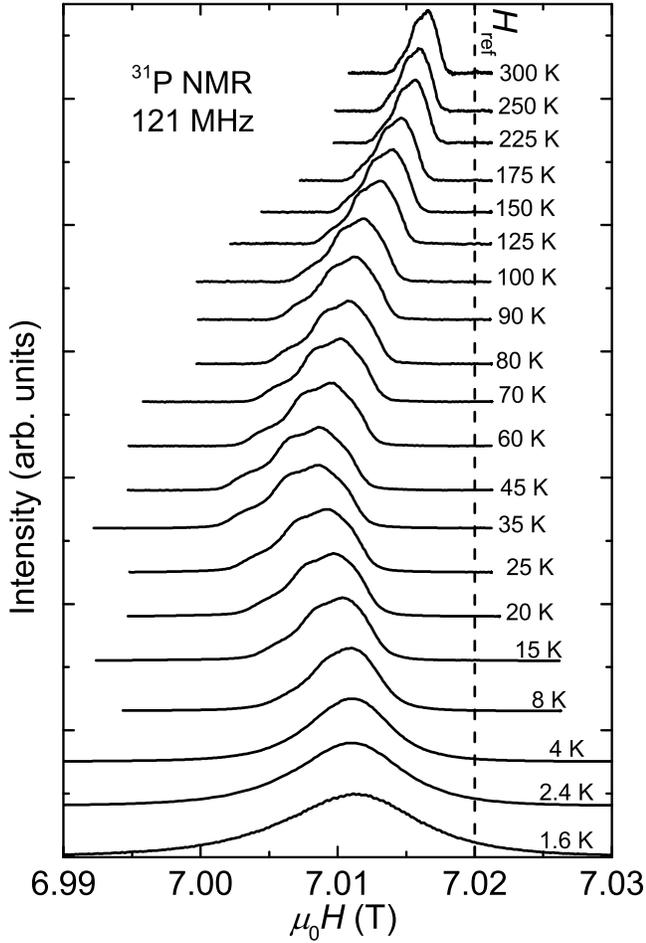}
	\caption {Field sweep $^{31}$P NMR spectra of KNaCuP$_2$O$_7$ at different temperatures measured in 121~MHz. The dashed line indicates the reference field position.}
	\label{Fig7}
\end{figure}
As presented in Fig.~\ref{Fig7}, we obtained a narrow and single spectral line at high temperatures, as expected for a $I = 1/2$ nucleus. The line shape is asymmetric and the central line position shifts with temperature. The asymmetric line shape reflects either asymmetry in the hyperfine coupling or anisotropic spin susceptibility. As the temperature is lowered, the line width also increases. Further, there are two in-equivalent P sites in the crystal structure and both of them are coupled to the Cu$^{2+}$ ions. Thus, our experimentally observed single spectral line in the whole measured temperature range implies that the local environment of both the P sites is nearly same. Indeed, a careful analysis of the crystal structure revels that the atomic positions of both the P sites are very close to each other. Further, no significant line broadening or change in line shape was observed down to 1.6~K, ruling out the low temperature magnetic LRO.

\subsubsection{$^{31}$P NMR Shift}
\begin{figure}
	\includegraphics[width=\linewidth]{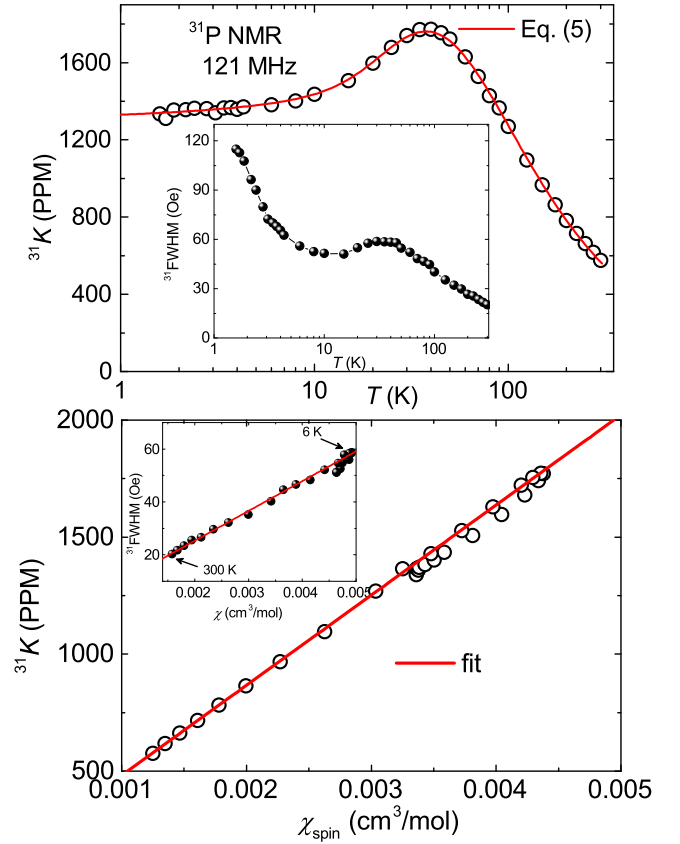}
	\caption {Upper panel: $^{31}$P NMR shift ($^{31}K$) vs temperature in 121~MHz. Solid line is the fit using Eq.~\eqref{K}. Inset: Full width at half maximum ($^{31}$FWHM) vs $T$. Lower panel: $^{31}K$ vs $\chi_{\rm spin}$ measured at $H = 1$~T in the $T$-range 2~K to 300~K. Solid line is a linear fit. Inset: $^{31}$FWHM vs $\chi$ and the solid line is a linear fit.}
	\label{Fig8}
\end{figure}
The temperature dependent NMR shift [$^{31}K(T)$] extracted from the central peak position is shown in Fig.~\ref{Fig8}. Similar to $\chi(T)$, $^{31}K(T)$ also passes through a broad maxima at around 40~K, footprint of the 1D short-range correlations. The noteworthy characteristic of $^{31}K(T)$ is that $^{31}K(T)$ has a great advantage over the bulk $\chi(T)$. At low temperature $\chi(T)$ shows a Curie-tail which originates mostly from either extrinsic paramagnetic impurities or defects in the powder sample. In contrast, NMR shift is completely insensitive to these contributions and probes only the intrinsic spin susceptibility, as $^{31}$P nucleus is coupled only to the Cu$^{2+}$ spins in the chain. Thus, the $^{31}K(T)$ data allow us to do a more accurate analysis of $\chi_{\rm spin}$ than $\chi(T)$. Moreover, the effect of impurity and defect contributions appears in the form of NMR line broadening. Therefore, the line width as a function of temperature should follow the bulk $\chi(T)$. One can expressed $^{31}K(T)$ in terms of $\chi_{\rm spin}(T)$ as 
\begin{equation}
\label{K}
^{31}K(T) = K_0 +\left(\frac{^{31}A_{\rm hf }}{N_{\rm A}\mu_{\rm B}}\right)\chi_{\rm spin}(T),
\end{equation}
where, $K_0$ is the temperature-independent chemical shift and $^{31}A_{\rm hf }$ is the average hyperfine coupling between $^{31}$P nucleus and Cu$^{2+}$ ions. The plot of $^{31}K$ versus $\chi_{\rm spin}$ with $T$ as an indirect variable is shown in the lower panel of Fig.~\ref{Fig8}. Here, $\chi_{\rm spin}$ at $H = 1$~T is taken from Fig.~\ref{Fig4}. The plot exhibits a nice straight line over the whole temperature range. From the slope of the linear fit, the total hyperfine coupling constant is calculated to be $^{31}A_{\rm hf} \simeq 2151.2$~Oe/$\mu_{\rm B}$.

In order to establish the spin-lattice and to extract the exchange coupling, $^{31}K(T)$ data were fitted using Eq.~\eqref{K}, taking the expression of $\chi_{\rm spin}(T)$ for a spin-$1/2$ uniform Heisenberg AF chain model.\cite{Johnston9558} It is apparent from Fig.~\ref{Fig8} that Eq.~\eqref{K} provides an excellent fit to the data in the entire temperature range 1.6~K$\leq T \leq 300$~K, unambiguously corroborating the 1D character of the spin-lattice. While fitting, the value of hyperfine coupling was kept fixed to $A_{\rm hf} \simeq 2151$~Oe/$\mu_{\rm B}$, obtained from the $^{31}K - \chi$ analysis. The obtained best fit parameters are $K_{0} \simeq 52.74$~ppm, $J/k_{\rm B} \simeq 58.7$~K, and $g \simeq 2.17$.

Theoretically, $\chi_{\rm spin}(T)$ or $K(T)$ for a spin-$1/2$ uniform HAF chain is predicted to show a logarithmic decrease $(\rm{ln}$$T^{-1}$) at low temperature ($T < 0.1 J/k_{\rm B}$) and reaches a finite value at $T=0$~K.\cite{Eggert332} The exact value of spin susceptibility at zero temperature can be estimated as $\chi_{\rm spin}(T = 0)=\frac {N_{\rm A}g^{2}\mu_{\rm B}^{2}}{J\pi^2}$.\cite{Johnston9558,GriffithsA768}
Experimentally, $\chi(T)$ and $^{17}$O $K(T)$ data of Sr$_2$CuO$_3$ and $^{31}$P $K(T)$ data of (Sr,Ba)$_2$Cu(PO$_4$)$_2$ and K$_2$CuP$_2$O$_7$, at very low temperatures are reported to show such a logarithmic decrease.\cite{Motoyama3212,Nath174436,Nath134451} For Sr$_2$CuO$_3$ with $J/k_{\rm B} \simeq 2200$~K, the decrease was observed at $T \simeq 0.01J/k_{\rm B}$ in $\chi(T)$\cite{Motoyama3212} and at $k_{\rm B}T/J\simeq 0.015$ in $K(T)$.\cite{Thurber247202} Similarly, for (Sr,Ba)$_2$Cu(PO$_4$)$_2$ ($J/k_{\rm B} \simeq 160$~K) and K$_2$CuP$_2$O$_7$ ($J/k_{\rm B} \simeq 141$~K) the decrease in $K(T)$ was observed below $T\simeq 0.003J/k_{\rm B}$ and $0.028J/k_{\rm B}$, respectively.\cite{Nath174436,Nath134451}
However, in KNaCuP$_2$O$_7$ $^{31}K(T)$ attains a finite value $\sim 1334$~PPM at 1.6~K, without any logarithmic decrease. Moreover, this value is found to be larger than the theoretically expected value $K_{\rm theo}(T = 0~{\rm K})=K_0 + \frac{A_{\rm hf}g^{2}\mu_{\rm B}}{J\pi^{2}} \simeq 1234$~PPM, taking $J/k_{\rm B} \simeq 58.7$~K, $^{31}A_{\rm hf} \simeq 2151$~Oe/$\mu_{\rm B}$, and $g=2.17$. In our case, the lowest measured temperature of 1.6~K corresponds to $\sim 0.03 J/k_{\rm B}$ only. This implies that one may needs to go further below 1.6~K inorder to see the low temperature decrease in $^{31}K(T)$.

The full width at half maximum ($^{31}$FWHM) of the $^{31}$P NMR spectra as a function of temperature is shown in the inset of the upper panel of Fig.~\ref{Fig8}. It displays a broad maximum at around 35~K and a Curie tail below 10~K, suggesting that $^{31}$FWHM traces the bulk $\chi(T)$, as expected. 

\subsubsection{$^{31}$P spin-lattice relaxation rate $^{31}1/T_1$}
\begin{figure}
	\includegraphics[width=\linewidth]{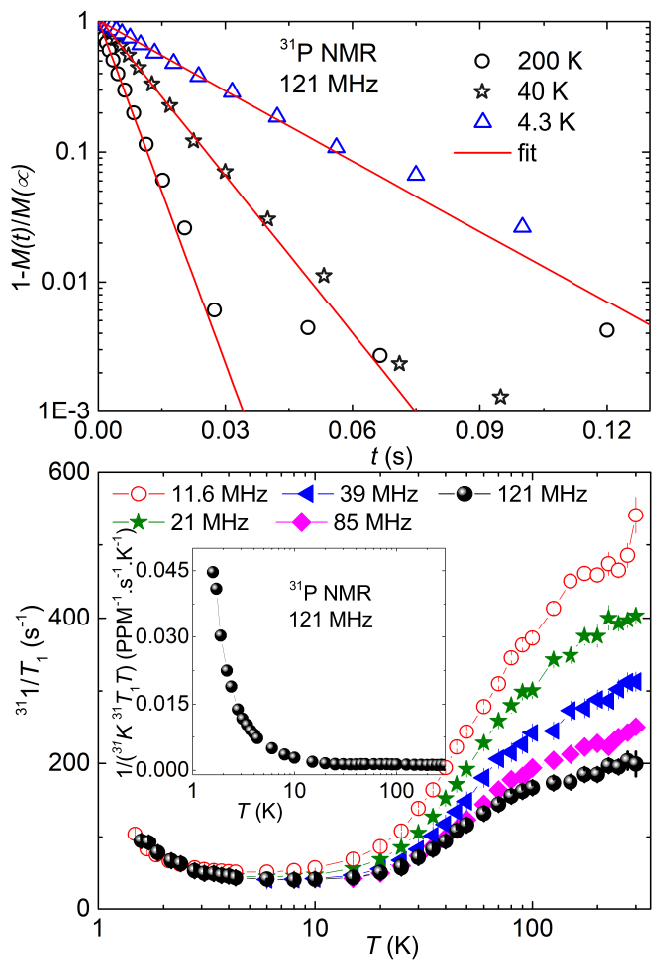}
	\caption {Upper panel: Longitudinal magnetization recovery curves at three selective temperatures measured on the $^{31}$P nuclei and the solid lines are fits using Eq.~\eqref{exp}. Lower panel: $^{31}$P NMR spin-lattice relaxation rate ($^{31}1/T_1$) as a function of temperature measured in different frequencies. The $x$-axis is shown in log scale in order to highlight the features in different temperature regimes. Inset: $1/(^{31}K ^{31}T_{1}T)$ vs $T$ for 121~MHz.}
	\label{Fig9}
\end{figure}
The $^{31}$P spin-lattice relaxation rate $^{31}1/T_{\rm 1}$ was measured at the field corresponding to the central peak position at each temperature. The longitudinal magnetization recoveries at three selected temperatures are shown in the upper panel of Fig.~\ref{Fig9}. As $^{31}$P is a $I = 1/2$ nucleus one can fit the recoveries by a single exponential function
\begin{equation}
1-\frac{M(t)}{M(\infty)}= Ae^{-t/T_{1}},
\label{exp}
\end{equation}
where $M(t)$ is the nuclear magnetization at a time $t$ after the saturation pulse and $M(\infty)$ is the equilibrium ($t \rightarrow \infty$) magnetization. Indeed, all the recovery curves could be fitted well by Eq.~\eqref{exp} (see upper panel of Fig.~\ref{Fig9}) and the curves show linearity over more than two decades when the $y$-axis is plotted in log scale. The extracted $^{31}1/T_1$ as a function of temperature measured at different frequencies are shown in the lower panel of Fig.~\ref{Fig9}. For the data at 121~MHz, $^{31}1/T_1$ is almost constant for $T > 90$~K which is typical due to the random movement of the paramagnetic moments.\cite{Moriya516} As the temperature is lowered further, $^{31}1/T_1$ decreases in a linear manner down to 20~K and then exhibits a temperature independent behaviour between 20~K and 4~K. At very low temperatures ($T < 4$~K), $^{31}1/T_1$ increases rapidly which indicates the slowing down of the fluctuating moments as the system approaches the magnetic LRO at $T_{\rm N}$. From the low temperature trend of $^{31}1/T_1$, the magnetic LRO is expected to set in at around $T_{\rm N} \sim 1$~K.

\subsubsection{$^{23}$Na NMR Spectra}
\begin{figure}
	\includegraphics[width=\linewidth]{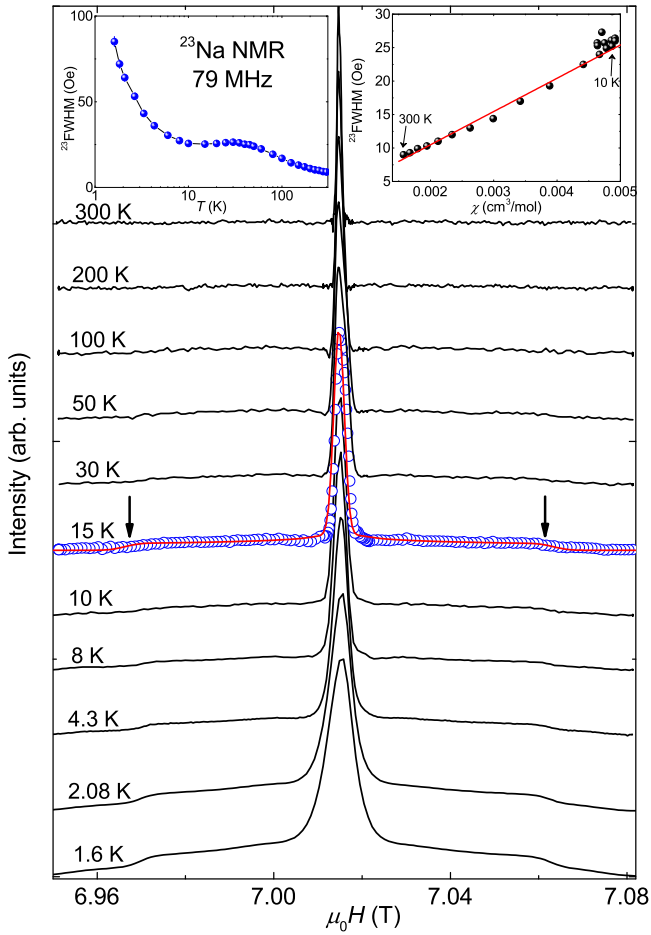}
	\caption {Field sweep $^{23}$Na NMR spectra of KNaCuP$_2$O$_7$ at different temperatures. The solid line is the fit of the spectrum at $T = 15$~K and the satellites are marked by arrows. Left inset: $^{23}$FWHM vs $T$. Right inset: $^{23}$FWHM vs $\chi$ and the solid line is a linear fit.}
	\label{Fig10}
\end{figure}
Since $^{23}$Na is quadrupolar nucleus with $I = 3/2$, the NMR line should have three lines. The central line corresponding to $I_z =+1/2\longleftrightarrow-1/2$ transition and two equally spaced satellite lines corresponding to $I_z =\pm3/2\longleftrightarrow\pm1/2 $ transitions on either side of the central line. The $^{23}$Na spectra as a function of temperature is presented in Fig.~\ref{Fig10}. At high temperatures, the line is very narrow and slightly asymmetric. As the temperature is lowered, the line width increases and two broad humps or satellites on both sides of the central line become prominent~\cite{Sebastian2021}.
However, the overall line shape remains invariant down to 1.6~K. Further, the position of the central line doesnot shift at all with temperature which confirms a weak hyperfine coupling of $^{23}$Na with the Cu$^{2+}$ ions due to negligible overlap of orbitals. This also justifies why the interchain interaction via Na is so weak.
The spectrum at $T = 15$~K could be fitted well with $K_{\rm iso} \simeq -60$~ppm (isotropic shift), $K_{\rm axial} \simeq 20$~ppm (axial shift), $K_{\rm aniso} \simeq 50$~ppm (anisotropic shift), $\eta = 0$ (asymmetry parameter), and $\nu_Q \simeq 0.57$~MHz (NQR frequency). The quadrupole frequency is almost temperature independent in the whole temperature range, which essentially excludes the possibility of any structural distortion in the studied compound. The $^{23}$FWHM with temperature, obtained from the fit is shown in the inset of Fig.~\ref{Fig10}. It passes through a broad maximum and then shows a low temperature Curie tail, identical to the bulk $\chi(T)$.

\subsubsection {$^{23}$Na spin-lattice relaxation rate $^{23}1/T_1$}
\begin{figure}
	\includegraphics{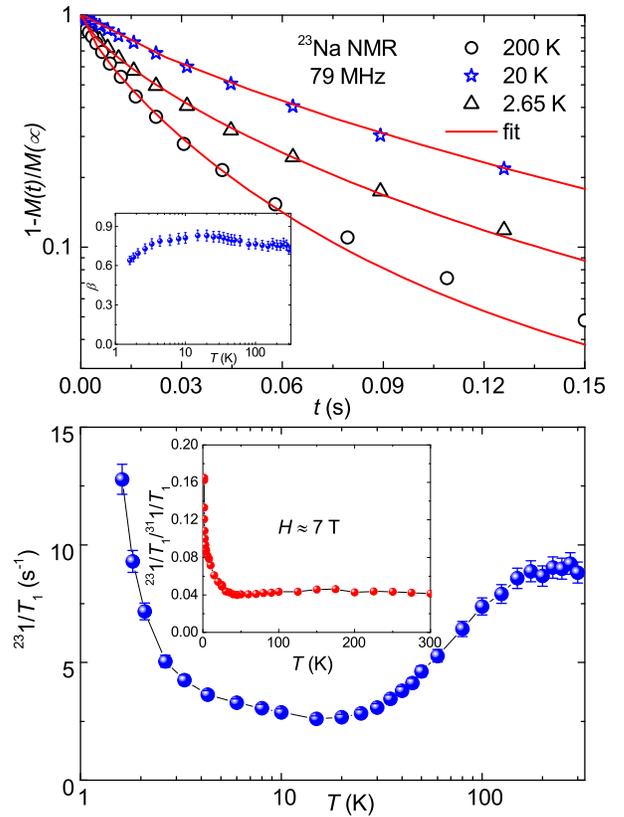}
	\caption{\label{Fig11} Longitudinal magnetization recovery curves at three selective temperatures measured on the $^{23}$Na nuclei and the solid lines are fits using Eq.~\eqref{Double_exp}. Inset: The exponent $\beta$ as a function of $T$. Lower panel: $^{23}1/T_{1}$ as a function of $T$. Inset: The ratio of relaxation rates $^{23}1/T_1$ and $^{31}1/T_1$ vs $T$ measured at $H \simeq 7$~T.}
\end{figure}
$^{23}1/T_1$ was measured by irradiating the central line of the $^{23}$Na spectra, choosing an appropriate pulse width. The recovery of the longitudinal magnetization was fitted well by the following 
double stretch exponential function~\cite{Gordon783,*Simmons1168}
\begin{equation}
\label{Double_exp}
1 - \frac{M(t)}{M(\infty)} = A[0.1~{\rm exp}(-t/T_1)^{\beta}+ 0.9~{\rm exp}(-6t/T_1)^{\beta}],
\end{equation}
relevent for the $^{23}$Na ($I=3/2$) nuclei. Here, $\beta$ is the stretch exponent. The upper panel of Fig.~\ref{Fig11} depicts recovery curves at three different temperatures. The obtained $^{23}1/T_1$ vs $T$ is shown in the lower panel of Fig.~\ref{Fig11}. The overall temperature dependence behaviour of $^{23}1/T_1$ is nearly identical to that observed for $^{31}1/T_1(T)$. For $T > 150$~K, $^{23}1/T_1$ is almost temperature independent. It decreases linearly below 150~K down to 30~K and remains constant between 30~K and 4~K. Below 4~K, $^{23}1/T_1$ shoots up and from the low-$T$ diverging trend one expects a peak at around $T_{\rm N} \simeq 1$~K, similar to $^{31}1/T_1$. The exponent $\beta$ as a function of $T$ is presented in the inset of the upper panel of Fig.~\ref{Fig11}. The absolute value of $\beta$ varies between 0.63 to 0.84. Such a reduced value of $\beta$ illustrates that there could be Na deficiency, as Na being the lightest element in the compound.

\subsection{Electronic Structure Calculations}
\begin{figure}
	\includegraphics [scale = 0.7] {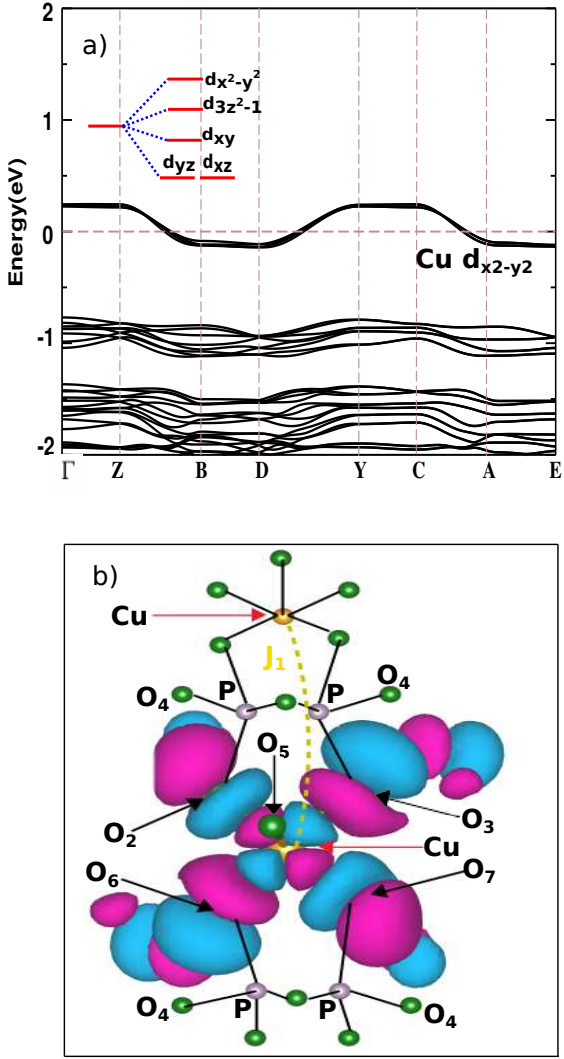}
	\caption{\label{Fig12} (a) Non-spin polarised band dispersion along various high symmetry directions. Inset shows the crystal field splitting. (b) Wannier function of effective Cu-$d_{x^{2}-y^{2}}$ orbital.}
\end{figure}
\label{sec:model}
\begin{table}
	\caption{\label{tab:exchange}
		Exchange parameters of KNaCuP$_2$O$_7$ obtained from DFT calculations: Cu--Cu distances $d$ (in\,\r A), electron hoppings $t_i$ (in\,meV), AFM contributions to the exchange $J_i^{\rm AFM}=4t_i^2/U_{\rm eff}$ (in\,K), and total exchange couplings $J_i$ (in\,K) from the LSDA+$U$ mapping procedure with $U_{\rm eff}=6.5$~eV.}
	\begin{ruledtabular}
		\begin{tabular}{cc@{\hspace{4em}}cc@{\hspace{4em}}r}
			& $d_{\rm Cu-Cu}$ & $t_i$ & $J_i^{\rm AFM}$ & $J_i$ \\\hline
			$J$      & 5.17 & 98 &     69      &  59  \\
			$J^{\prime}$ & 5.67 & 2.17  &   $\sim 0.1$   & $\sim 0.1$  \\
			$J^{\prime \prime}$ & 5.77 & 0.14 & $\sim 0.1$  & $\sim 0.1$  \\
		\end{tabular}
	\end{ruledtabular}
\end{table}
First principles electronic structure calculations in the framework of DFT have been carried out to identify the dominant exchange paths, the various exchange couplings, and the resulting spin model. In order to get insights on the electronic structure of KNaCuP$_{2}$O$_{7}$, we have started with the non-spin polarized calculations [see Fig.~\ref{Fig12}(a)]. Our calculations revealed O-$p$ states are completely occupied while K, Na, and P states are empty, consistent with the nominal ionic formula K$^{1+}$ Na$^{1+}$ Cu$^{2+}$ P$_{2}^{5+}$ O$_{7}^{2-}$, indicating Cu is in 3$d^9$ configuration. As a consequence, the Fermi level is dominated by four Cu-$d$ bands arising from the four Cu atoms in the four formula unit cell of KNaCuP$_{2}$O$_{7}$ [see Fig.~\ref{Fig12}(a)]. In the local frame of reference, {\em i.e.} assuming that the Cu atom is residing at the origin and choosing the $z$-axis along the long Cu-apical O bond, $x$ and $y$ axes along Cu-O bonds in the basal plane, we find that these bands at the Fermi level are predominantly of Cu-$d_{x^{2}-y^{2}}$ character. The band structure shows strong dispersion parallel to the chain direction Z-B and D-Y but is nearly dispersionless perpendicular to the direction of the chains, indicating strong 1D character of this system.
	
In order to evaluate the Cu intersite exchange strengths, we have calculated exchange interactions using the 'four state' method\cite{Xiang224429} based on the total energy of the system with few collinear spin alignments. If the magnetism in the system is fully described by the Heisenberg Hamiltonian [$\textbf{H}= \sum_{ij} J_{ij} S_i \cdot S_j$], the energy for such a spin pair can be written as follows:
	\begin{equation}\label{ham}
		E= J_{12}\textbf{S}_{1}.\textbf{S}_{2} + \textbf{S}_{1}.\textbf{h}_1 + \textbf{S}_{2}.\textbf{h}_2 +E_{all} + E_{0},
	\end{equation}
where, we consider the exchange interaction $J_{12}$ between spins at site 1 and 2. $\textbf{h}_1= \sum_{i\neq1,2} J_{1i}\textbf{S}_{i}, \textbf{h}_2= \sum_{i\neq1,2} J_{2i}\textbf{S}_{i}, E_{\rm all}= \sum_{i\neq1,2} J_{ij}\textbf{S}_{i}.\textbf{S}_{j}$, and $E_{0}$ contains all other non-magnetic energy contributions. The second (third) term in Eq.~\eqref{ham} corresponds to the coupling of the spin 1(2) with all other spins in the unit cell excluding spin 2(1). $E_{\rm all}$ takes into account the exchange couplings between all spins in the unit cell except from spins 1 and 2. The exchange interaction strength between sites 1 and 2 is obtained by considering four collinear spin states $(i)~1_{\uparrow},2_{\uparrow}$, $(ii)1_{\uparrow},2_{\downarrow}$, $(iii)1_{\downarrow},2_{\uparrow}$, and $(iv)1_{\downarrow},2_{\downarrow}$ as
\begin{equation}
\label{spin_state} J_{12}= \frac{E_{\uparrow \uparrow}+E_{\downarrow \downarrow}-E_{\uparrow \downarrow}-E_{\downarrow \uparrow}}{4S^2}.
\end{equation}
The first (second) suffix of energy ($E$) represents the spin state of site 1(2). The estimated exchange interactions along with the corresponding Cu-Cu distances [as depicted in Fig.~\ref{Fig1}(b)] are tabulated in Table~\ref{tab:exchange}. The NN exchange interaction is found to be the strongest one and AFM ($J/k_{\rm B} = 59$~K) which is in excellent agreement with the experiment.
The other exchange interactions $J^{\prime}$ and $J^{\prime\prime}$ are abysmally small (0.1~K) and are AFM adding inter-chain frustration to the system. Further, the calculated mean-field Curie-Weiss temperature $\theta_{\rm CW}= 29$~K, compares well with the experiment\cite{Bhowal075110}.

Finally, the $Cu-d_{x^{2}-y^{2}}$ Wannier function has been plotted for KNaCuP$_{2}$O$_{7}$ in Fig.~\ref{Fig12}(b). The tails of the $Cu-d_{x^{2}-y^{2}}$ orbital are shaped according to the $O p_{x} / p_{y}$ orbitals such that $Cu-d_{x^{2}-y^{2}}$ forms strong pd$\sigma$ anti-bonds with the $O p_{x} / p_{y}$ tails in the basal plane. We see that the Cu–Cu hopping primarily proceeds via the oxygens. The  dominant intrachain AFM exchange interaction $J$ is mediated via Cu-O-P-O-Cu path, while the other interchain exchage interactions are mediated via the long Cu-O bond along the apical oxygen (2.32~$\AA$), thereby rendering them to be weak.

\section {Discussion}
We have demonstrated that KNaCuP$_2$O$_7$ is a good example of an 1D spin-$1/2$ uniform HAF. KNaCuP$_2$O$_7$ formally belongs to the family of $A_2$CuP$_2$O$_7$ ($A$ = Na, Li, and K) compounds, although they have different crystal structures. KNaCuP$_2$O$_7$ has monoclinic structure with space group $P2_1/n$ in contrast to monoclinic unit cell with space group $C2/c$ for (Na,Li)$_2$CuP$_2$O$_7$ and orthorhombic unit cell with space group $Pbnm$ for K$_2$CuP$_2$O$_7$.\cite{Erragh23,*Gopalakrishna1171,*Elmaadi13} For (Na,Li)$_2$CuP$_2$O$_7$, a slightly distorted CuO$_4$ plaquettes are corner shared with PO$_4$ tetrahedra making spin chains with an intra-chain exchange coupling $J/k_{\rm B} \simeq 28$~K and magnetic LRO at $T_{\rm N} \simeq 5$~K.\cite{Nath4285,Lebernegg174436} Here, the neighbouring plaquettes are tilted toward each other by an angle of about $70^{\degree}$ and $90^{\degree}$ for Na and Li compounds, respectively resulting in a buckling of the spin chains. This modulation in spin chains is responsible for a weaker intra-chain coupling and magnetic LRO at a relatively high temperature. On the other hand, for K$_2$CuP$_2$O$_7$, the arrangement of CuO$_4$ plaquettes are more planar and the chains are strictly straight which give rise to pronounced 1D magnetism with a larger intra-chain coupling $J/k_{\rm B}\simeq141$~K and without any magnetic LRO down to 2~K.\cite{Nath134451}
For KNaCuP$_2$O$_7$, though the CuO$_4$ plaquettes are arranged in the same plane, similar to K$_2$CuP$_2$O$_7$ but they are more distorted with four different Cu-O bonds distances ($\sim 1.932 - 1.987$~\AA). Further, the Cu-Cu inter-chain distances are slightly reduced for KNaCuP$_2$O$_7$ ($\sim 5.6767 - 7.01$~\AA) compared to K$_2$CuP$_2$O$_7$ ($\sim 5.879 - 7.388$~\AA). Because of the difference in the structural arrangements, the intra-chain (NN) exchange coupling of KNaCuP$_2$O$_7$ ($J/k_{\rm B}\simeq58.7$~K) has an intermediate value between K$_2$CuP$_2$O$_7$ and (Na,Li)$_2$CuP$_2$O$_7$.

Further, the inter-chain couplings which are unavoidable in experimental compounds, drive the system into a LRO state at a finite temperature. However, when the inter-chain couplings form a frustrated network, the ground state is modified significantly and in many cases forbid the compound going to a LRO state. The magnetic LRO at a very low temperature ($T_{\rm N} \simeq 1$~K) in KNaCuP$_2$O$_7$ evidences extremely weak as well as frustrated inter-chain exchange couplings. With this value of $T_{\rm N}$, the compound exhibits one-dimensionality over a large temperature range $k_{\rm B}T_{\rm N}/J \simeq 1.7 \times 10^{-2}$. One can tentatively estimate the average interchain coupling ($J^{\prime}$) of a quasi-1D HAF chain by putting the appropriate values of $J$ and $T_{\rm N}$ in the simple expression obtained from the mean-field approximation\cite{Irkhin6757,Johannes174435}
\begin{equation}
	\label{TN_3D model}
	J^{\prime}/k_{\rm B} =\frac{3.046 T_{\rm N}}{z k_{\rm AF}\sqrt{ln\left(\frac{5.8 J}{k_{\rm B}T_{\rm N}}\right)+0.5~ {ln}~{ ln}\left(\frac{5.8 J}{k_{\rm B}T_{\rm N}}\right)}}.
\end{equation}
Here, $k_{\rm AF}$ represents the AF wave vector and $z=6$ is the number of nearest neighbour spin chains. Numerical calculations for a 3D model yield $k_{\rm AF}\simeq0.70$. For KNaCuP$_2$O$_7$, using $J/k_{\rm B} \simeq 58.7$~K and $T_{\rm N} \simeq 1$~K, the average inter-chain coupling is estimated to be $J^{\prime}/k_{\rm B} (= J^{\prime \prime}/k_{\rm B}) \simeq 0.28$~K. This value is of the same order of magnitude as that obtained from the electronic structure calculations.



Spin-lattice relaxation rate, $1/T_1$, provides useful information on spin dynamics or dynamic susceptibility of a spin system. It helps to access the low-energy spin excitations by probing the nearly zero-energy limit (in the momentum space) of the local spin-spin correlation function~\cite{Moriya23}.
Quite generally, $\frac{1}{T_{1}T}$ is written in terms of the dynamic susceptibility $\chi_{M}(\vec{q},\omega_{0})$ as\cite{Moriya516}
\begin{equation}
	\frac{1}{T_{1}T} = \frac{2\gamma_{N}^{2}k_{B}}{N_{\rm A}^{2}}
	\sum\limits_{\vec{q}}\mid A(\vec{q})\mid
	^{2}\frac{\chi^{''}_{M}(\vec{q},\omega_{0})}{\omega_0},
	\label{t1form}
\end{equation}
where the sum is over the wave vector $\vec{q}$ within the first Brillouin zone, $A(\vec{q})$ is the form-factor of the hyperfine interaction, and $\chi^{''}_{M}(\vec{q},\omega _{0})$ is the imaginary part of the dynamic susceptibility at the nuclear Larmor frequency $\omega _0$.
Thus, $1/T_1$ has contributions from both uniform ($q=0$) and staggered ($q = \pm \pi/a$) spin fluctuations. For 1D spin-$1/2$ chains, theory predicts that the uniform component leads to $1/T_1 \propto T$ while the staggered component gives $1/T_1=$~constant~\cite{Sachdev13006,SandvikR9831}. Typically, $q = \pm \pi/a$ and $q = 0$ components dominate the $1/T_1$ data in the low temperature ($T << J/k_{B}$) and high temperature ($T \sim J/k_{B}$) regimes, respectively~\cite{Nath174436}. Thus, the experimentally observed linear decrease and temperature independent behaviour of $1/T_1$ in the intermediate temperature ranges reflect the dominance of $q=0$ and $q = \pm \pi/a$ contributions, respectively.

As discussed earlier, $^{31}$P is located symmetrically between two adjacent Cu$^{2+}$ ions along the chain. Similarly, $^{23}$Na is coupled, though weakly, to four Cu$^{2+}$ ions from three neighboring chains. Therefore, the staggered components of the hyperfine fields from the neighbouring Cu$^{2+}$ ions are expected to be cancelled out at both the $^{31}$P and $^{23}$Na sites. Accordingly, one should be able to probe the low energy spin excitations corresponding to the $q = 0$ mode separately from the staggered $q = \pm \pi/a$ mode. However, in our case, there is still significant contribution from $q = \pm \pi/a$ which dominates the low temperature $1/T_1$ data. One possible source of the remnant staggered fluctuations could be the unequal hyperfine couplings arising due to low symmetry of the crystal structure. Further, the linear and constant temperature regimes are found to be different for $^{31}1/T_{\rm 1}$ and $^{23}1/T_{\rm 1}$ which is likely due to subtle difference in the hyperfine form factors associated with the $^{31}$P and $^{23}$Na nuclei.
In Eq.~\eqref{t1form} for $q = 0$ and $\omega_0 = 0$, the real component of $\chi_{\rm M}^{\prime}(q,\omega_0)$ represents to the static susceptibility $\chi$ (or $K$). Therefore, $1/(\chi T_{1}T)$ should be temperature independent. As shown in the inset of the lower panel of Fig.~\ref{Fig9}, $1/(^{31}K ^{31}T_1T)$ indeed demonstrates the dominant contribution of $\chi$ to $1/(^{31}T_1T)$. However, a slight increase in $1/(^{31}K^{31}T_1T)$ below $\sim 5$~K indicates the growth of AF correlations with decreasing $T$. Moreover, when ratio of $^{23}1/T_1$ at 79~MHz ($H \simeq 7.0147$~T) and $^{31}1/T_1$ at 121~MHz ($H \simeq 7.0203$~T) is plotted against temperature (see, inset of the lower panel of Fig.~\ref{Fig11}), it results an almost constant value above $\sim 40$~K and then increases rapidly towards low temperatures.


\begin{figure}
	\includegraphics{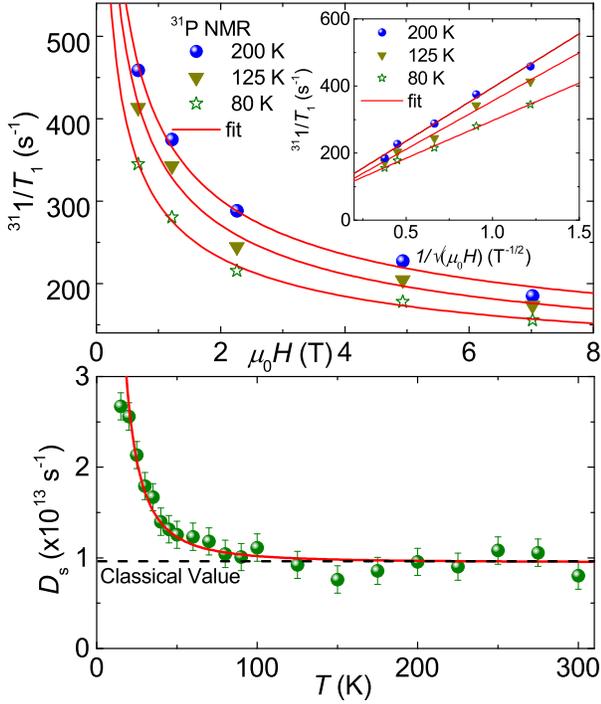}
	\caption{\label{Fig13} Upper panel: $^{31}$P NMR spin-lattice relaxation rate ($^{31}1/T_1$) as a function of applied magnetic field ($\mu_0H$) at $T= 80$~K, 125~K, and 200~K. The solid lines are the fits using $1/T_1=a+b/\sqrt{\mu_0H}$. Inset: $^{31}1/T_1$ vs $1/\sqrt{\mu_0H}$. Lower panel: Temperature dependence of $D_{\rm s}$ deduced from $^{31}1/T_1$. The solid line is the fit using $D_{\rm s} \sim 1/T^2$. The classically expected value at high temperatures is also shown as a dashed line.}
\end{figure}
In order to detect the effect of external magnetic field on the spin dynamics, we have measured $^{31}1/T_1$ vs $T$ at different frequencies/fields.
As seen in the lower panel of Fig.~\ref{Fig9}, $^{31}1/T_1$ shows a strong frequency dependency in the high temperature regime and the absolute value of $^{31}1/T_1$ decreases with increase in frequency. This difference is narrowed down as the temperature is lowered and below about 20~K, the data sets in different frequencies overlap with each other. It is established that the long wavelength ($q \sim 0$) spin fluctuations in a Heisenberg magnet often show diffusive dynamics. In 1D spin chains, such spin diffusion leads to a $1/\sqrt{H}$ field dependence of $^{31}1/T_1$~\cite{Borsa2215,Takigawa2173}.
Thus, the strong field dependency of $^{31}1/T_1$ at high temperatures appears to be due to the effect of spin diffusion where long wavelength $q = 0$ fluctuations dominate. Moreover, the weak field dependency of $^{31}1/T_1$ at low temperatures also reflects that the relaxation is dominated by the staggered ($q=\pm \pi/a$) fluctuations below 20~K.

The contribution of spin-diffusion to $1/T_1$ can be written as\cite{Hone965,Takigawa4612,Thurber247202}
\begin{equation}
	\frac{1}{T_{1}^{\rm sd}T} =\frac{A_{\rm hf}^{2}(q=0)\gamma_{\rm n}^{2} k_{B}\chi(T, q=0)}{\mu_{\rm B}^{2}\sqrt{2g\mu_{\rm B}D_{\rm s}H/\hbar}},
	\label{t1sd}
\end{equation}
where, $D_{\rm s}$ is the spin-diffusion constant. Thus, the slope of the linear $^{31}1/T_1$ vs $1/\sqrt{H}$ plot at a fixed temperature should yield $D_{\rm s}$. In the upper panel of Fig.~\ref{Fig13}, we have plotted $^{31}1/T_1$ vs $H$ for three different temperatures ($T= 80$~K, 125~K, and 200~K) which are fitted by $1/T_{1} = a+b/\sqrt{\mu_{\rm 0}H}$, where $a$ and $b$ are the constants. To magnify the linear behaviour, $^{31}1/T_1$ is plotted against $1/\sqrt{\mu_{\rm 0}H}$ in the inset of the upper panel of Fig.~\ref{Fig13}.
Using the value of $\chi(T)$ obtained from the NMR shift measurement and the slope ($b$) in Eq.~\eqref{t1sd}, the diffusion constant at each temperature is determined.
The temperature dependence of $D_{\rm s}$ deduced from $^{31}1/T_1$ is presented in the lower panel of Fig.~\ref{Fig13}. It increases moderately with decreasing temperature, as expected in the region dominated by the $q = 0$ fluctuations. The value of $D_{\rm s}$ in high temperatures ($T>100$~K) is of the same order as the classically expected value, $D_{\rm s} = (J/\hbar)\sqrt{2\pi S(S+1)/3} = 9.64\times 10^{12}$~sec$^{-1}$~\cite{Hone965}. This is indeed consistent with the previous reports on other Heisenberg spin chain compounds~\cite{Boucher4098,Takigawa2173,Takigawa4612,Fujiwara11945}. Further, the temperature dependent $D_{\rm s}$ could be fitted by $D_{\rm s} \sim 1/T^2$, similar to $^{17}$O NMR in Sr$_2$CuO$_3$~\cite{Thurber247202}. However, it is not clear whether such a behaviour of $D_{\rm s}(T)$ can be accounted for by the 1D spin-$1/2$ chain model.

\section {Conclusion}
Our results demonstrate that KNaCuP$_2$O$_7$ is an excellent 1D spin-$1/2$ HAF model system with nearest-neighbor only exchange. The magnetic susceptibility, magnetization isotherm, and $^{31}$P NMR shift data show good agreement with the theoretical predictions for 1D spin-$1/2$ HAF chain with intra-chain coupling $J/k_{\rm B}\simeq 58.7$~K. The value of intra-chain coupling is further confirmed from the complementary electronic structure calculations and the subsequent QMC simulations. From the $^{31}K$ vs $\chi_{\rm spin}$ plot, the hyperfine coupling of $^{31}$P with Cu$^{2+}$ ion is estimated to be $^{31}A_{\rm hf} \simeq 2151.2$~Oe/$\mu_{\rm B}$. The presence of magnetic LRO at a very low temperature provides evidence of an extremely weak and frustrated inter-chain couplings and one-dimensionality over a large temperature range $k_{\rm B}T_{\rm N}/J \simeq 1.7 \times 10^{-2}$. The moderate value of the exchange coupling allowed us to access the spin excitations of the spin-$1/2$ Heisenberg chain at both low and high temperature limits. The change of slope in $^{31}1/T_1(T)$ and $^{23}1/T_1(T)$ at around $T \sim 20-30$~K explain the crossover regime of the dominant contributions from the uniform ($q = 0$) and staggered ($q = \pm \pi/a$) spin fluctuations. Our results also established that the dynamic spin susceptibility has a strong diffusive contribution at high temperatures. However, the nature of the temperature dependent diffusion constant $D_{\rm s}$ is not yet understood.

\section {Acknowledgments}
Authors acknowledge I. Dasgupta for discussions regarding the theoretical work.
SG and RN would like to acknowledge BRNS, India for financial support bearing sanction No.37(3)/14/26/2017-BRNS. Work at the Ames Laboratory was
supported by the U.S. Department of Energy, Office of
Science, Basic Energy Sciences, Materials Sciences, and
Engineering Division. The Ames Laboratory is operated
for the U.S. Department of Energy by Iowa State University under Contract No. DEAC02-07CH11358. AG thanks SERB, India (Project No. EMR/2016/005925) and SM thanks CSIR, India for fellowship.


%

\end{document}